\documentclass[pra,aps,twocolumn,superscriptaddress,showpacs,tightenlines,amsmath]{revtex4}

\usepackage{epsfig}

\begin{document}
\title{Exploiting entanglement in communication channels with correlated noise}
\author{Jonathan Ball}
\affiliation{Centre for Quantum Computation, Clarendon Laboratory,
University of Oxford, Oxford OX1 3PU, United Kingdom}

\author{Andrzej Dragan}
\affiliation{Instytut Fizyki Teoretycznej, Uniwersytet Warszawski,
Ho\.{z}a 69, 00-681 Warszawa, Poland}

\author{Konrad Banaszek}
\affiliation{Centre for Quantum Computation, Clarendon Laboratory,
University of Oxford, Oxford OX1 3PU, United Kingdom}

\begin{abstract}
We develop a model for a noisy communication channel in which the
noise affecting consecutive transmissions is correlated. This
model is motivated by fluctuating birefringence of fiber optic
links. We analyze the role of entanglement of the input states in
optimizing the classical capacity of such a channel. Assuming a
general form of an ensemble for two consecutive transmissions, we
derive tight bounds on the classical channel capacity depending on
whether the input states used for communication are separable or
entangled across different temporal slots. This result
demonstrates that by an appropriate choice, the channel capacity
may be notably enhanced by exploiting entanglement.
\end{abstract}


\maketitle
\section{Introduction}

Entanglement is a fragile feature of composite quantum systems
that can easily diminish by uncontrollable interactions with the
environment. At the same time however carefully crafted entangled
states can protect quantum coherence from the deleterious effects
of those random interactions. This idea underlies the principles
of quantum error correcting codes that strengthen the optimism
regarding the feasibility of implementing in practice complex
quantum information processing tasks \cite{NielsenChuang}.

In this paper we demonstrate how quantum entanglement can help in
the task of classical communication. To this end, we develop a
simple model of a noisy communication channel, where the noise
affecting consecutive transmissions is correlated. Within this
model, we derive bounds on the classical channel capacity assuming
either separable or entangled input states, and we show that using
collective entangled states of transmitted particles leads to an
enhanced capacity of the channel.

The motivation for our model comes from classical fiber optic
communications \cite{GhatakThyagarajan}. In practice, light
transmitted through a fiber optic link undergoes a random change
of polarization induced by the birefringence of the fiber. The
fiber birefringence usually fluctuates depending on the
environmental conditions such as temperature and mechanical
strain. At first sight, this makes the polarization degree of
freedom unsuitable for encoding information, as the input
polarization state gets scrambled on average to a completely mixed
state. However, the birefringence fluctuations have a certain time
constant which means that the transformation of the polarization
state, though random, remains nearly the same on short time
scales. Consider now sending a pair of photons whose temporal
separation lies well within this time scale. Although the
polarization state of each one of the photons when looked at
separately becomes randomized, certain properties of the joint
state remain preserved. For example, this is the case of the
relative polarization of the second photon with respect to the
first one. We can therefore try to decode from the output whether
the input polarizations were mutually parallel or orthogonal. This
property cannot be determined perfectly, as in general we cannot
tell whether two general quantum states are identical or
orthogonal if we do not know anything else about them
\cite{KashKentPRA02}, but even the ability of providing a partial
answer establishes correlations between the channel input and
output that can be used to encode information into the
polarization degree of freedom. The situation becomes even more
interesting when we allow for entangled quantum states. Then the
singlet polarization state of the two photons, when sent as the
input, remains invariant under such perfectly correlated
depolarization, and it can be discriminated unambiguously against
the triplet subspace. Therefore we can encode one bit of
information into the polarization state of two photons by sending
either a singlet state or any of the triplet states. We shall see
that these simple observations will also emerge from our general
analysis of the channel capacity.

The first example of entanglement-enhanced information
transmission over a quantum channel with correlated noise has been
recently analyzed by Macchiavello and Palma \cite{MaccPalmPRA02}.
Our model assumes a different form of correlations, and its high
degree of symmetries has allowed us to perform optimization of the
channel capacity over arbitrary input ensembles. Although we
analyze only zero- and one-photon signals, we define the action of
the channel in terms of the transformations of the bosonic
annihilation operators, which sets up a framework for possible
generalizations, such as use of multiphoton signals. This
application of entanglement in classical communication is a
distinct problem from entanglement-assisted classical capacity of
noisy quantum channels studied by Bennett {\em et al.} in
Ref.~\cite{BennShorPRL99}, where it has been shown that prior
entanglement shared between sender and receiver can increase the
classical capacity. We also note that the non-zero time constant
of phase and polarization fluctuations can be used in robust
protocols for long-haul quantum key distribution
\cite{WaltAbouPRL03,BoilGootXXX03}.

Before passing on to a detailed discussion of the problem in the
subsequent sections, let us introduce some basic notation. The
action of a channel is described by a completely positive map
\cite{HausJozsPRA96} that we will denote by $\Lambda(\cdot)$. The
sender selects messages from an input ensemble $\{ p_i,
\hat{\varrho}_i \}$, where $p_i$ is the probability of sending the
state $\hat{\varrho}_i$ through the channel. The capacity of the
channel is a function of the mutual information between the input
ensemble and measurement outcomes at the receiving stations: it
characterizes the strength of correlations between these two that
are preserved by the channel. The mutual information itself
involves a specific measurement scheme; however, it has a very
useful upper bound in the form of the Holevo quantity that depends
only on the output ensemble of states $\{ p_i,
\Lambda(\hat{\varrho}_i) \}$ emerging from the channel
\cite{Holevo}:
\begin{equation}
\chi = S\left(\sum_i p_i \Lambda(\hat{\varrho}_i) \right) -
\sum_{i} p_i S(\Lambda(\hat{\varrho}_i))
\end{equation}
where $S$ is the von Neumann entropy
$S(\hat{\varrho})=-\text{Tr}(\hat{\varrho}\log_2\hat{\varrho})$.
As we will see, in our model the Holevo quantity will provide a
tight bound on the mutual information that could be achieved in
practice using a simple measurement scheme. The classical channel
capacity is obtained by assuming arbitrarily long sequences of
possibly entangled input systems, and calculating the average
capacity per single use of the channel. In our analysis, we will
perform a restricted optimization by considering only two
consecutive uses of the channel.

\section{Channel decomposition}

We will start our discussion by proving a rather general lemma
about channels that can be decomposed into a direct sum of maps
acting on subspaces of the Hilbert space of the input systems. In
physical terms, such channels remove quantum coherence between the
components of the input state that belong to different subspaces,
by zeroing the respective off-diagonal blocks of the density
matrix characterizing the input state. This lemma will greatly
simplify our further calculations.

{\em Lemma 1:} Suppose that we can decompose the Hilbert space
${\cal H}$ of the system into a direct sum of subspaces
\begin{equation}
{\cal H} = \bigoplus_{k} {\cal H}^{(k)}
\end{equation}
such that for an arbitrary input state $\hat{\varrho}$ the state
emerging from the channel $\Lambda(\hat{\varrho})$ can be
represented as
\begin{equation}
\Lambda({\hat{\varrho}}) = \bigoplus_{k}
\Lambda^{(k)}(\hat{\varrho}^{(k)})
\end{equation}
where $\hat{\varrho}^{(k)} = {\hat{\varrho}}|_{{\cal H}^{(k)}}$ is
the input state $\hat{\varrho}$ truncated to the subspace ${\cal
H}^{(k)}$, and each $\Lambda^{(k)}$ is a certain trace-preserving
completely positive map acting in the corresponding subspace
${\cal H}^{(k)}$. Then the optimal channel capacity can be
attained with an ensemble in which each state belongs to one of
the subspaces ${\cal H}^{(k)}$.

{\em Proof:} Indeed, suppose that there is a state $\hat{\varrho}$
that does not satisfy the above condition, i.e.\ it is defined on
more that one subspace ${\cal H}^{(k)}$. We can replace it by a
sub-ensemble
$\{\text{Tr}(\hat{\varrho}^{(k)});\hat{\varrho}^{(k)}/
\text{Tr}(\hat{\varrho}^{(k)})\}$, obtained by truncating the
state $\hat{\varrho}$ to the subspaces ${\cal H}^{(k)}$ and
normalizing the resulting density matrices. In other words,
whenever the sender is supposed to transmit $\hat{\varrho}$, she
replaces it by one of the normalized truncated states
$\hat{\varrho}^{(k)}/\text{Tr}(\hat{\varrho}^{(k)})$ with the
corresponding probability $\text{Tr}(\hat{\varrho}^{(k)})$. It is
straightforward to verify that the average state obtained from
such a subensemble is identical with $\Lambda(\hat{\varrho})$.

The above observation has a useful consequence when optimizing the
Holevo bound on channel capacity. If the input ensemble is of the
form discussed above, then it can be split into subensembles of
states that belong to separate subspaces ${\cal H}^{(k)}$, with
the probability distributions normalized to one within each
subensemble, and $p_k$ denoting the probability of sending a state
from the $k$th subensemble. It is then easy to check that the
Holevo quantity is given by the following expression:
\begin{equation}
\label{Eq:chi=sumpkchik} \chi=\sum_{k} p_k \chi^{(k)}-\sum_{k} p_k
\log_2 p_k,
\end{equation}
where $\chi^{(k)}$ is the Holevo quantity for the $k$th
subensemble. Therefore, the maximization of the Holevo quantity
can be performed in two steps. The first one is the optimization
of each of $\chi^{(k)}$ separately, assuming an input ensemble
restricted to the subspace ${\cal H}^{(k)}$. The second step
consists of optimizing the probability distribution $p_k$ with the
normalization constraint $\sum_k p_k = 1$, and it can be performed
explicitly using the method of Lagrange multipliers. Indeed, if we
denote the Lagrange multiplier as $\lambda$, then differentiation
over $p_l$ yields:
\begin{equation}
0 = \frac{\partial}{\partial p_l} \left( \chi - \lambda \sum_{k}
p_k \right) =  \chi^{(l)} - \log_2 p_l - \frac{1}{\ln 2} -
\lambda.
\end{equation}
This formula allows us to express the probabilities $p_l$ in terms
of the Lagrange multiplier $\lambda$ as:
\begin{equation}
\label{Eq:pl} p_l = 2^{\chi^{(l)}-1/\ln 2 - \lambda},
\end{equation}
and furthermore summation over $l$ and using the fact that $\sum_l
p_l =1$ gives the value of the Lagrange multiplier as:
\begin{equation}
\label{Eq:mu} \lambda = \log_2 \left( \sum_l 2^{\chi^{(l)}}
\right) - \frac{1}{\ln 2}.
\end{equation}
Finally, inserting Eqs.~(\ref{Eq:pl}) and (\ref{Eq:mu}) into
Eq.~(\ref{Eq:chi=sumpkchik}) yields the maximum value of the
Holevo quantity equal to:
\begin{equation}
\label{Eq:ChiSum} \chi=\log_2 \left( \sum_k 2^{\chi^{(k)}}
\right).
\end{equation}
We will later find this expression useful in calculating the
channel capacity in our model. The physical reason for this is
that we will be able to decompose the set of states used for
communication into subensembles with a fixed number of photons,
and then optimize the Holevo quantity separately in each subspace.

\section{Depolarization model}

Let us now introduce a mathematical model for the random
transformation of polarization during transmission through the
channel. A general linear transformation between two annihilation
operators corresponding to a pair of orthogonal modes is given by
$2\times 2$ unitary matrices \cite{CampSalePRA89} that form the
Lie group U(2). In situations when only the relative phase between
the two polarization modes is relevant, the overall phase of the
transformation can be assumed to be fixed, which reduces the group
of transformations to SU(2). However, in our case the overall
phase shift can vary between the consecutive temporal slots, and
therefore we need to keep it as an independent parameter. We note
that any U(2) matrix can be mapped onto a rotation in the three
dimensional physical space.  Such a rotation describes the
corresponding transformation of the Poincar\'{e} sphere used to
represent the polarization state of light in classical optics
\cite{BornWolf}. We will label elements of  U(2) as ${\bf\Omega}$
and use a dot to denote the multiplication within the group. The
U(2) group has a natural invariant integration measure which we
assume is normalized to one $\int \text{d}{\bf\Omega} = 1$. This
measure defines a uniformly randomized distribution of
polarization transformations that scrambles an arbitrary input
polarization to a completely mixed one.

\begin{figure}
\epsfig{width=3in,file=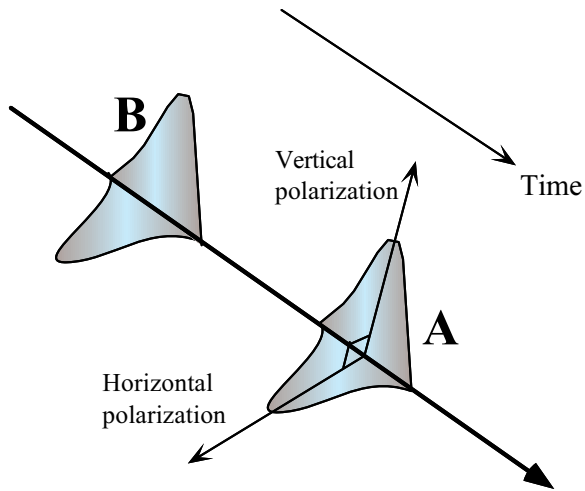} \caption{Representation of
two consecutive temporal slots labelled by $A$ and $B$. The
Hilbert space of each slot is spanned in our model by three
states: the zero-photon state $|0\rangle$ and two mutually
orthogonal polarization states denoted by
$|\leftrightarrow\rangle$ and $|\updownarrow\rangle$.}
\label{fig:tempslots}
\end{figure}

Suppose now that two consecutive temporal slots labelled by $A$
and $B$, each comprising two orthogonal polarizations, are
occupied by a joint state of radiation $\hat{\varrho}_{AB}$, as
shown schematically in Fig.~\ref{fig:tempslots}. We will assume
that the polarization transformation ${\bf\Omega}_A$ affecting the
slot $A$ is completely random, but that the transformation
${\bf\Omega}_B$ is correlated with the first one through a
conditional probability distribution $p({\bf\Omega}_B |
{\bf\Omega}_A)$. The resulting transformation of the joint
two-slot state is therefore given by the following completely
positive map:
\begin{eqnarray}
\Lambda(\hat{\varrho}_{AB}) & = & \int d{\bf\Omega}_A \int
d{\bf\Omega}_B \, p({\bf\Omega}_B | {\bf\Omega}_A)
\nonumber \\
& & \times \hat{U}({\bf\Omega}_A) \otimes \hat{U}({\bf\Omega}_B)
\, \hat{\varrho}_{AB} \, \hat{U}^\dagger({\bf\Omega}_A) \otimes
\hat{U}^\dagger({\bf\Omega}_B).
\nonumber \\
& &
\end{eqnarray}
Here $\hat{U}({\bf\Omega})$ is a unitary matrix acting in the
Hilbert space of one of the slots that represents the polarization
transformation ${\bf\Omega}$. We will now assume that the
conditional probability $p({\bf\Omega}_B | {\bf\Omega}_A)$ depends
only on the relative transformation between the slots $A$ and $B$
and that it can consequently be represented as $p({\bf\Omega}_B |
{\bf\Omega}_A) = p({\bf\Omega}_B \cdot {\bf\Omega}_A^{-1})$. In
such a case, we can substitute the integration variables in the
second integral according to ${\bf\Omega}_B={\bf\Omega}' \cdot
{\bf\Omega}_A$, and make use of the invariance of the integration
measure $\text{d}{\bf\Omega}_B = \text{d} {\bf\Omega}'$. This
procedure shows that the map $\Lambda$ can be represented as a
composition of two maps: $\Lambda = (\hat{\openone} \otimes
\Lambda_{\text{dep}}) \circ \Lambda_{\text{perf}}$. The first one
of them,  $\Lambda_{\text{perf}}$, acts on both the temporal slots
and it depolarizes them in exactly the same way:
\begin{equation}
\label{Eq:Lambdaperf} \Lambda_{\text{perf}}(\hat{\varrho}_{AB}) =
\int d{\bf\Omega}\, \hat{U}({\bf\Omega}) \otimes
\hat{U}({\bf\Omega}) \, \hat{\varrho}_{AB} \,
\hat{U}^\dagger({\bf\Omega}) \otimes \hat{U}^\dagger({\bf\Omega})
\end{equation}
The second map, $\Lambda_{\text{dep}}$, acts only on the slot $B$,
and it introduces additional depolarization relative to the slot
$A$ according to the probability distribution $p({\bf\Omega}')$:
\begin{equation}
\Lambda_{\text{dep}} (\hat{\varrho}_B) = \int d{\bf\Omega}' \,
p({\bf\Omega}') \hat{U}({\bf\Omega}') \hat{\varrho}_B
\hat{U}^\dagger({\bf\Omega}').
\end{equation}
We will assume later that the distribution $p({\bf\Omega}')$ has
sufficient symmetry to describe the action of the map
$\Lambda_{\text{dep}}$ in the relevant Hilbert space with the help
of two simple parameters.

We now introduce a further simplification by imposing a condition
that each temporal slot may contain at most one photon. Therefore
the relevant Hilbert space for each slot is spanned by three
states: the zero-photon state $|0\rangle$, and horizontally and
vertically polarized one-photon states $|\leftrightarrow\rangle$
and $|\updownarrow\rangle$. We can conveniently write the explicit
form of the unitary transformation $\hat{U}({\bf\Omega})$ using
the irreducible unitary representations of the group SU(2). We
will denote by $\hat{\cal D}^{j}({\bf\Omega})$ a
$(2j+1)\times(2j+1)$ matrix that is a $(2j+1)$-dimensional
representation of an SU(2) element obtained from ${\bf\Omega}$ by
fixing the overall phase factor to one. These matrices are well
known in the quantum theory of angular momentum as describing
transformations of a spin-$j$ particle under the rotation group
\cite{BrinkSatchler}. We will also denote by $\alpha({\bf\Omega})$
the overall phase of the element ${\bf\Omega}$. Then the unitary
transformation of the input state corresponding to the
polarization rotation ${\bf\Omega}$ is given by the matrix:
\begin{equation}
\hat{U}({\bf\Omega}) = \left(
\begin{array}{cc}
\hat{\cal D}^{0}({\bf\Omega}) &
\begin{array}{cc} 0 & 0 \end{array}
\\
\begin{array}{c} 0 \\ 0
\end{array}
& e^{i\alpha({\bf\Omega})}\hat{\cal D}^{1/2}({\bf\Omega})
\end{array}
\right).
\end{equation}
In this formula, the one-dimensional representation $\hat{\cal
D}^{0}({\bf\Omega})$ is identically equal to one, and
$e^{i\alpha({\bf\Omega})}\hat{\cal D}^{1/2}({\bf\Omega})$ is a
$2\times 2$ unitary matrix itself; however, we will keep this more
general notation in order to be able to use results from the
theory of group representations. In particular, the following
property of the rotation matrix elements will allow us to evaluate
directly a number of expressions:
\begin{equation}
\label{Eq:IntDD} \int\text{d}{\bf\Omega} [ {\cal
D}^{j}_{mn}({\bf\Omega})]^{\ast} {\cal D}^{j'}_{m'n'}({\bf\Omega})
= \frac{1}{2j+1} \delta_{jj'} \delta_{mm'} \delta_{nn'}.
\end{equation}

The action of the map $\Lambda_{\text{perf}}$ on a joint two-slot
state can be analyzed most easily if we decompose the complete
Hilbert space into a direct sum of subspaces with a fixed number
of photons: ${\cal H} = {\cal H}^{(0)} \oplus {\cal H}^{(1)}
\oplus {\cal H}^{(2)}$, where the upper index labels the number of
photons. The zero-photon subspace is spanned by a single state
$|0_A 0_B \rangle$. The one-photon space has a basis formed by
four vectors: $|\leftrightarrow_A 0_B\rangle$, $|\updownarrow_A
0_B\rangle$, $|0_A \leftrightarrow_B \rangle$, and $|0_A
\updownarrow_B \rangle$. Finally, in the two-photon subspace
${\cal H}^{(2)}$ we will introduce a basis that consists of the
singlet state $|\Psi_-\rangle = (|\leftrightarrow_A
\updownarrow_B\rangle - |\updownarrow_A
\leftrightarrow_B\rangle)/\sqrt{2}$ and the three triplet states
$|\leftrightarrow_A \leftrightarrow_B\rangle$, $|\Psi_+\rangle =
(|\leftrightarrow_A \updownarrow_B\rangle + |\updownarrow_A
\leftrightarrow_B\rangle)/\sqrt{2}$, and $|\updownarrow\rangle_A
\updownarrow\rangle_B$. The reason for this choice is that then
the action of the tensor product $\hat{\cal D}^{1/2}({\bf\Omega})
\otimes \hat{\cal D}^{1/2}({\bf\Omega})$ on a two-photon state can
be decomposed into the sum: $\hat{\cal D}^{1/2}({\bf\Omega})
\otimes \hat{\cal D}^{1/2}({\bf\Omega}) = \hat{\cal
D}^{0}({\bf\Omega}) \oplus \hat{\cal D}^{1}({\bf\Omega})$ where
$\hat{\cal D}^{0}({\bf\Omega})$ acts on the singlet state
$|\Psi_-\rangle$, and $\hat{\cal D}^{1}({\bf\Omega})$ is a
three-dimensional matrix acting in the triplet subspace. Using our
decomposition of the complete Hilbert space, the action of the
tensor product $\hat{U}({\bf\Omega}) \otimes \hat{U}({\bf\Omega})$
on a general two-slot state in the basis specified above is given
by:
\begin{widetext}
\begin{equation}
\hat{U}({\bf\Omega}) \otimes \hat{U}({\bf\Omega}) = \hat{\cal
D}^{0}({\bf\Omega}) \oplus e^{i\alpha({\bf\Omega})} \left(
\begin{array}{cc}
\hat{\cal D}^{1/2}({\bf\Omega}) & \begin{array}{cc} 0 & 0 \\ 0 & 0 \end{array} \\
\begin{array}{cc} 0 & 0 \\ 0 & 0 \end{array} & \hat{\cal D}^{1/2}({\bf\Omega})
\end{array}
\right) \oplus e^{2i\alpha({\bf\Omega})} \left(
\begin{array}{cc}
\hat{\cal D}^{0}({\bf\Omega}) & \begin{array}{ccc} 0 & 0 & 0 \end{array} \\
\begin{array}{c}
0 \\ 0 \\ 0
\end{array}
& \hat{\cal D}^{1}({\bf\Omega})
\end{array}
\right)
\end{equation}
\end{widetext}
If we now insert this fomula into Eq.~(\ref{Eq:Lambdaperf}), it
can be easily seen that the invariant integration over the overall
phase factor $\alpha({\bf\Omega})$ kills all the off-block
diagonal elements of the density matrix that link different
subspaces ${\cal H}^{(k)}$. In other words, all the coherence
between states with different photon numbers is completely removed
by the phase fluctuations. Furthermore, the operation
$\Lambda_{\text{dep}}$, acting only on the second slot, does not
mix subspaces with different photon numbers. Therefore the
conditions of our lemma are satisfied and we can consider only
states with a definite number of photons as elements of the input
ensemble. Thus we need to calculate are three corresponding Holevo
quantities $\chi^{(0)}$, $\chi^{(1)}$, and $\chi^{(2)}$ that can
be combined into a Holevo bound for the overall channel capacity
according to Eq.~(\ref{Eq:ChiSum}). This calculation forms the
contents of the next section.

\section{Channel capacity}
\label{Sec:ChannelCapacity}

The communication capacity $\chi^{(0)}$ of the zero-photon
subspace itself ${\cal H}^{(0)}$ is naturally zero, as we have
only a single state  $|0_A 0_B\rangle$ at our disposal. This state
can of course be used as an element of a larger ensemble thus
contributing to the overall capacity. This fact is reflected in
the form of Eq.~(\ref{Eq:ChiSum}), where $\chi^{(0)}=0$ indeed
does increase the total value of $\chi$.

\subsection{One-photon subspace}

A less trivial problem to calculate is the capacity of the
one-photon subspace. If we assume a normalized input state
$\hat{\varrho}_{\text{in}}$ from the subspace ${\cal H}^{(1)}$,
then the action of the channel $\Lambda_{\text{perf}}$ restricted
to this subspace is given by:
\begin{equation}
\label{Eq:Lambda(1)perfrhoin}
\Lambda^{(1)}_{\text{perf}}(\hat{\varrho}_{\text{in}}) =
\frac{1}{2}\left(
\begin{array}{cccc}
a & 0 & b & 0 \\
0 & a & 0 & b \\
b^\ast & 0 & 1-a & 0 \\
0 & b^\ast & 0 & 1-a
\end{array}
\right)
\end{equation}
where the parameters $a$ and $b$ are defined in terms if the input
density matrix as:
\begin{eqnarray}
a & = & \langle \leftrightarrow_A 0_B | \hat{\varrho}_{\text{in}}
| \leftrightarrow_A 0_B \rangle + \langle \updownarrow_A 0_B |
\hat{\varrho}_{\text{in}} | \updownarrow_A 0_B \rangle
\nonumber \\
b & = & \langle \leftrightarrow_A 0_B | \hat{\varrho}_{\text{in}}
| 0_A \leftrightarrow_B \rangle + \langle \updownarrow_A 0_B |
\hat{\varrho}_{\text{in}} | 0_A \updownarrow_B \rangle
\end{eqnarray}
For the form of  density matrix given in
Eq.~(\ref{Eq:Lambda(1)perfrhoin}), the depolarizing channel
$\Lambda_{\text{dep}}$ affects only the off-diagonal elements $b$
and $b^\ast$. We will assume that the symmetry of the distribution
$p({\bf\Omega}')$ is such that the effect of
$\Lambda_{\text{dep}}$ is a rescaling of these elements by a real
parameter $\eta'$ bounded between $0$ and $1$. It is now easy to
check that the entropy of the one-photon state emerging from the
channel can be written as
\begin{equation}
S(\Lambda(\hat{\varrho}_{\text{in}})) = 1 + S ( \left(
\begin{array}{cc} a & \eta' b \\ \eta' b^\ast & 1-a \end{array}
\right) )
\end{equation}
where the $2\times 2$ matrix appearing in the second term can be
interpreted as a state of a qubit. Therefore, the second term is
bounded by $0$ and $1$, and consequently $1 \leq
S(\Lambda(\hat{\varrho}_{\text{in}})) \leq 2$. It is a
straightforward observation that the Holevo quantity is bound from
above by the difference between the maximum and the minimum
possible entropies of states emerging from the channel. Therefore
we obtain that $\chi^{(1)} \leq 1$. This inequality can be
saturated simply by taking a one-photon state confined either to
the first or to the second temporal slot, with an arbitrary
polarization. Thus, the channel capacity is not enhanced in the
one-photon sector.

\subsection{Two-photon subspace}

The most interesting regime is when both the temporal slots are
occupied by photons. As we will see  below, in this case quantum
correlations can then enhance the capacity of the channel. If we
take a normalized input state $\hat{\varrho}_{\text{in}}$ from the
two-photon subspace ${\cal H}^{(2)}$, then the map
$\Lambda_{\text{perf}}$ produces a Werner state \cite{WernPRA89}:
\begin{equation}
\Lambda_{\text{perf}}^{(2)}(\hat{\varrho}_{\text{in}}) =
\hat{W}_c,
\end{equation}
where we have introduced the following notation:
\begin{equation} \hat{W}_c = - c |\Psi_-\rangle \langle
\Psi_- | + (1+c) \frac{\hat{\openone}}{4}
\end{equation}
and we will use for $c$ the name of the Werner parameter of the
input state $\hat{\varrho}_{\text{in}}$, defined as:
\begin{equation}
\label{Eq:c} c=\frac{1}{3}-\frac{4}{3}\langle \Psi_-
|\hat{\varrho}_{\text{in}} |\Psi_-\rangle.
\end{equation}
This result, derived previously in Ref.~\cite{WernPRA89}, can be
verified independently using the property given in
Eq.~(\ref{Eq:IntDD}).

The second operation affecting the input state is the partially
depolarizing channel $\hat{\openone} \otimes
\Lambda_{\text{dep}}$. We will assume that the action of the map
$\Lambda_{\text{dep}}$ acting on the photon in the second temporal
slot is simply isotropic depolarization shrinking the length of
the Bloch vector by a factor $\eta$ satisfying $0\leq\eta\leq1$.
Such an operation preserves the Werner form of the transmitted
state, and its only effect is the multiplication of the parameter
$c$ by the factor $\eta$. Thus, the state emerging from the
channel is given by:
\begin{equation}
\Lambda^{(2)}(\hat{\varrho}_{\text{in}}) = \hat{W}_{\eta c}
\end{equation}
with the parameter $c$ defined by the input state
$\hat{\varrho}_{\text{in}}$ according to Eq.~(\ref{Eq:c}).

At this point the possibility of enhanced communication capacity
by exploiting entanglement manifests itself. The difference
between the separable and entangled alphabets can be seen by
comparing the allowed ranges of the parameter $c$. The positivity
of the input density matrix $\hat{\varrho}_{\text{in}}$ requires
that
\begin{equation}
\label{Eq:EntangledContraint} {-1 \le c \le 1/3}
\end{equation}
and this is the only condition if we consider the most general,
possibly entangled input states. However, if the input states are
restricted to \textit{separable} ones, then as shown by Horodeccy
\cite{HoroHoroPRA96}, the allowed range for the parameter $c$ is
reduced to
\begin{equation}
\label{Eq:SeparableConstraint} {-1/3\le c \le 1/3}.
\end{equation}
This limitation will underlie the reduced channel capacity in the
case of separable states.

As the two-photon states emerging from the channel are fully
characterized by the Werner parameters of the respective input
states, optimization of the Holevo quantity can be carried out
over the ensemble $\{q_j ; c_j \}$ of the probabilities $q_j$ of
sending the $j$th state with the Werner parameter equal to $-c_j$.
The output states emerging from the channel is therefore given by
an ensemble of Werner states $\{q_j ; \hat{W}_{\eta c_j} \}$.
Because a statistical mixture of Werner states is also a Werner
state with the average parameter:\begin{equation} \sum_j q_j
\hat{W}_{\eta c_j} = \hat{W}_{\sum_j q_j \eta c_j},
\end{equation}
the Holevo quantity can be expressed with the help of a single
real-valued function $f(c)$:
\begin{eqnarray}
\chi^{(2)} & = & S\left( \sum_j q_j \hat{W}_{\eta c_j} \right) -
\sum_j q_j S(\hat{W}_{\eta c_j} )
\nonumber \\
\label{Eq:Chi2} & = & f\left(\sum_j q_j \eta c_j \right) - \sum_j
q_j f(\eta c_j)
\end{eqnarray}
where the explicit form of the function $f(c)$ is given by:
\begin{equation}
\label{Eq:f(c)} f(c) = 2 -\frac{3}{4}(1+c)\log_2 (1+c) -
\frac{1}{4} (1-3c) \log_2 (1-3c).
\end{equation}
The optimization of the Holevo quantity, which in principle needs
to be performed over an arbitrarily large input ensemble of
permitted quantum states, can be greatly simplified using the
following observation.

{\it Lemma 2}: Let $f(\gamma)$ be a concave function defined on a
closed interval $[\alpha,\beta]$, and let $q_j$ be a probability
distribution for a set $\gamma_j$ of real numbers taken from the
range $\alpha \le \gamma_j \le \beta$. Then the following
inequality holds:
\begin{eqnarray}
\lefteqn{ f\left( \sum_j q_j \gamma_j  \right) - \sum_j q_j
f(\gamma_j) }
& & \nonumber \\
\label{Eq:fsumqgamma} & \le & \sup_{\alpha \le \gamma \le \beta}
\left( f(\gamma) - \frac{\beta-\gamma}{\beta - \alpha} f(\alpha) -
\frac{\gamma-\alpha}{\beta - \alpha} f(\beta) \right).
\end{eqnarray}

{\em Proof:} The concavity of the function $f(c)$ implies that for
every $j$ we have:
\begin{equation}
f(\gamma_j) \ge \frac{\beta-\gamma_j}{\beta - \alpha} f(\alpha) +
\frac{\gamma_j-\alpha}{\beta- \alpha} f(\beta).
\end{equation}
If we now multiply the above equation by $-q_j$, perform the
summation over $j$, and add a term $\sum_{j} f(q_j \gamma_j)$ to
both sides of the equation, we will obtain an inequality whose
left hand side is identical with that of
Eq.~(\ref{Eq:fsumqgamma}), and the right hand side is exactly the
argument of the supremum for $\gamma = \sum_j q_j \gamma_j$.
Obviously, this value of $\gamma$ lies between $\alpha$ and
$\beta$, and consequently the supremum may only exceed the value
obtained from this calculation. This confirms that
Eq.~(\ref{Eq:fsumqgamma}) is indeed satisfied.

The above lemma reduces the whole problem of optimizing the Holevo
bound to maximizing a one-parameter real-valued function that is
the argument of the supremum on the right hand side of
Eq.~(\ref{Eq:fsumqgamma}). Inserting the explicit form of the
function $f(\gamma)$ given in Eq.~(\ref{Eq:f(c)}) and
differentiating the resulting expression over $\gamma$ shows that
the supremum in the right hand side of Eq.~(\ref{Eq:fsumqgamma})
is attained for
\begin{equation}
\label{Eq:TurningPoint}
\gamma_{\text{opt}}=\frac{1-2^{4\mu/3}}{3+2^{4\mu/3}}
\end{equation}
where $\mu=[f(\beta)-f(\alpha)]/(\beta-\alpha)$.

As we have seen, the permitted range of the parameters $c_j$
characterizing the states belonging to the input ensemble  depends
on whether we allow most general, possibly entangled states, or
rather restrict the input to separable states only. If we assume
that this range spans from $c_{\text{min}}$ to $c_{\text{max}}$:
\begin{equation}
c_{\text{min}} \le c_j \le c_{\text{max}}
\end{equation}
then we can easily apply Lemma 2 to the expression of the Holevo
quantity $\chi^{(2)}$ in terms of the function $f(c)$ that has
been given in the second line of Eq.~(\ref{Eq:Chi2}). Taking
$\alpha=\eta c_{\text{min}}$ and $\beta = \eta c_{\text{max}}$ and
using the explicit value of the turning point derived in
Eq.~(\ref{Eq:TurningPoint}) yields the following bound:
\begin{equation}
\label{Eq:chi(2)le} \chi^{(2)} \le \log_2(3+2^{4\mu/3})-f(\eta
c_{\text{min}}) +\mu(\eta c_{\text{min}}-1/3)-2
\end{equation}
where $\mu$ is given in terms of the input ensemble
characteristics as:
\begin{equation}
\mu=\frac{f(\eta c_{\text{max}})-f(\eta
c_{\text{min}})}{\eta(c_{\text{max}}-c_{\text{min}})}.
\end{equation}
We will analyze in detail numerical values of the channel capacity
in the next section. Before doing so, we will close this section
by describing a simple intuitive picture of Lemma~2 that gives an
additional insight into the form of the input ensemble.

\begin{figure}
\epsfig{width=3in,file=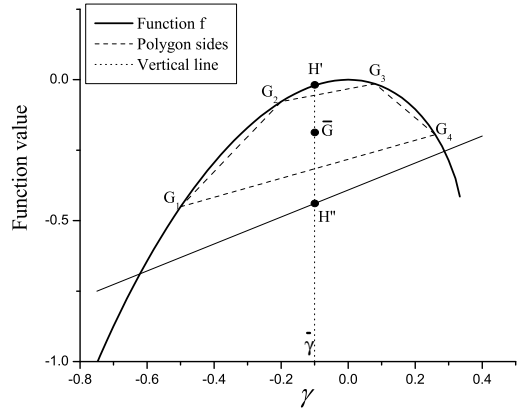} \caption{The graphical
representation the maximization procedure for the two-photon
subspace. The set of points $G_j$ corresponds to the output
ensemble. The difference $f(\gamma)-g(\gamma)$ over $\gamma$ needs
to be maximized over the interval $[\alpha,\beta]$.}
\label{fig:graphill}
\end{figure}

\subsection{Graphical interpretation}

The result of Lemma 2 can be visualized using the following
geometrical reasoning depicted in Fig.~\ref{fig:graphill}.
Consider a graph of the function $f(\gamma)$ versus its argument
$\gamma$. The numbers $\gamma_j$ and the corresponding values of
the function $f$ are given by a set of points
$G_j=(\gamma_j,f(\gamma_j))$ in the plane of the graph. The
probability distribution $q_j$ for the arguments $\gamma_j$
defines an average
\begin{equation}
\bar{G}=\left(\sum_j q_j \gamma_j, \sum_j q_j f(\gamma_j) \right)
\end{equation}
that can be interpreted as a center of gravity for the system of
points $G_j$ that have been assigned respective masses $q_j$.
Obviously, if the probability distribution is arbitrary, then this
average can lie anywhere within the convex polygon spanned by the
points $G_j$. Since the function $f$ is strictly concave over the
range considered, the whole polygon lies within the area bounded
by the graph of the function $f(\gamma)$ on one side, and a
straight line connecting the points $(\alpha,f(\alpha))$ and
$(\beta,f(\beta))$ on the other side. This straight line is given
by a function $g$ defined as:
\begin{equation}
g(\gamma)=\frac{\beta-\gamma}{\beta - \alpha} f(\alpha) +
\frac{\gamma-\alpha}{\beta - \alpha} f(\beta).
\end{equation}

The left hand side of Eq.~(\ref{Eq:fsumqgamma}) is now given by
the length of a vertical line connecting $\bar{G}$ with the point
$H'= (\bar{\gamma}, f(\bar{\gamma}))$ on the graph of the function
$f(\gamma)$, where $\bar{\gamma}=\sum_j q_j \gamma_j$. Clearly,
the line $\bar{G}H'$ will be always equal in length or shorter
than the line $H'H''$ where the point $H''= (\bar{\gamma},
g(\bar{\gamma}))$ lies on the graph of the function $g(\gamma)$.
Furthermore, in order to find the maximum possible length of the
line $H'H''$, it is clear from this geometric construction that we
need to maximize the difference $f(\gamma)-g(\gamma)$ over
$\gamma$ belonging to the interval $[\alpha,\beta]$. This
procedure is expressed explicitly in the right hand side of
Eq.~(\ref{Eq:fsumqgamma}) and the parameter $\mu$ introduced in
the previous subsection is simply the gradient of the function
$g(\gamma)$.

It is clearly seen from this geometric construction that enlarging
the interval $[\alpha,\beta]$ can only increase the value of the
upper bound given in Eq.~(\ref{Eq:fsumqgamma}). This implies two
rather straightforward observations. First, the use of entangled
states should give a larger capacity compared to separable states.
Secondly, a lower value of the parameter $\eta$ meaning weaker
correlations between consecutive polarization rotations results in
a decreased channel capacity.

\begin{figure}
\epsfig{width=3in,file=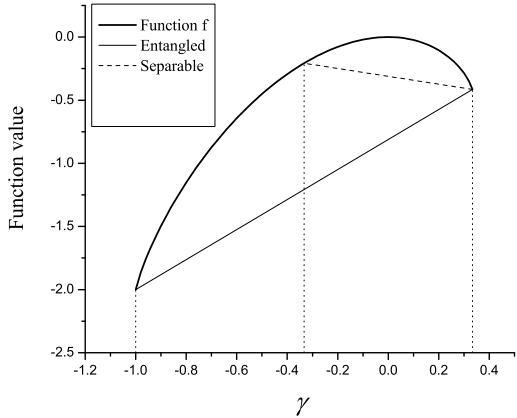} \caption{Depiction of the optimal
ensembles that maximize the Holevo quantity in both the general
entangled case and the restricted separable case, for perfectly
correlated noise ($\eta=1$). It is sufficient to take only
two-element ensembles with the extreme points of the allowed
interval. For general entangled states the interval is $[-1,1/3]$
whereas for the separable case the interval is reduced to
$[-1/3,1/3]$} \label{fig:ens}
\end{figure}

The graphical construction presented above also gives a simple
recipe for constructing an output ensemble that saturates the
bound on the Holevo quantity. It is sufficient to take a
two-element ensemble with the extreme points of the allowed
interval as the parameters of the Werner states emerging from the
channel: $\alpha=\eta c_{\text{min}}$ and $\beta=\eta
c_{\text{max}}$. The optimal probabilities of using the two states
need to be selected in such a way that the weighted sum of the
points corresponding to these states gives the point
$\gamma_{\text{opt}}$ maximizing the difference
$f(\gamma)-g(\gamma)$. Explicitly, these probabilities are
respectively given by $(\beta-\gamma_{\text{opt}})/(\beta-\alpha)$
and $(\gamma_{\text{opt}}-\alpha)/(\beta-\alpha)$. The actual
graph of the function $f(\gamma)$ with the permitted ranges of the
Werner parameter for perfectly correlated noise and entangled and
separable inputs is shown in Fig.~\ref{fig:ens}.

\section{Attainability and implementation}

The Holevo quantity $\chi$ is only an upper bound on the channel
capacity and therefore is not necessarily attainable. Users of a
communication channel need two relevant pieces of information. The
first one is the optimal form of the input ensemble that should be
used by the sender. The second one is a measurement scheme that
should be employed at the output of the channel in order to
optimize the capacity.

Let us start by summarizing the results of the preceding section
and specifying the input ensemble implied by these considerations.
We have seen that in the zero- and one-photon subspaces the
channel capacity cannot be enhanced by exploiting the polarization
degree of freedom. Therefore as the elements of the input ensemble
we can take for example states $|0_A 0_B \rangle$,
$|\updownarrow_A 0_B \rangle$, and $|0_A \updownarrow_B \rangle$,
where for concreteness we have fixed the polarization of
single-photon states to vertical. The polarization degree of
freedom starts to play a nontrivial role when both the temporal
slots are occupied by photons. In this subspace, we need to select
two input states characterized by the Werner parameters that are
as distant as it is allowed by the constraints on the input
ensemble. If we restrict ourselves to separable states, then
according to Eq.~(\ref{Eq:SeparableConstraint}) we need to take
one separable state with $c_{\text{min}}=-1/3$ and another one
with $c_{\text{max}}=1/3$. It is easy to verify using
Eq.~(\ref{Eq:c}) that the pair of separable states satisfying this
condition can be taken as $|\updownarrow_A \leftrightarrow_B
\rangle$ and $|\updownarrow_A \updownarrow_B \rangle$. We thus see
that in agreement with the simple picture developed in the
introduction to this paper, the relevant quantity is the relative
polarization of the photons occupying consecutive slots. If we
allow for entangled input, then the lower limit for the Werner
parameters of the input states shifts down to $c_{\text{min}}=-1$.
This value can be of course attained by taking the singlet state
$|\Psi_-\rangle$ itself as one element of the input ensemble, and
any state with $c_{\text{max}}=1/3$, for example again
$|\updownarrow_A \updownarrow_B \rangle$ as the second one.

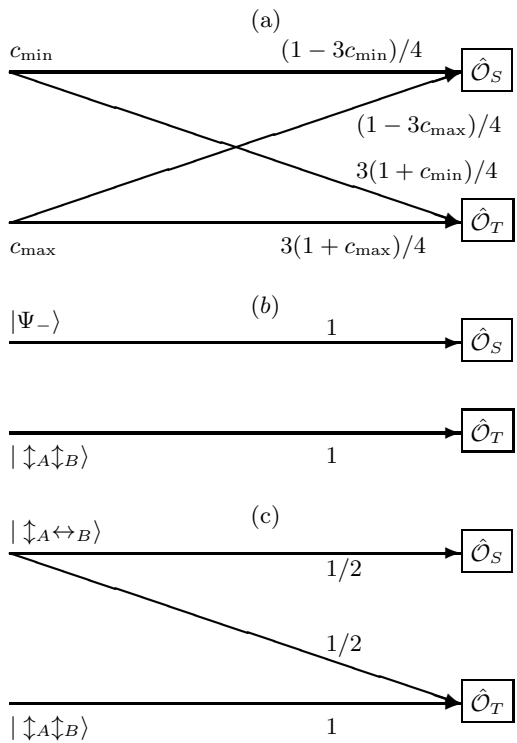
\begin{figure}
\setlength{\unitlength}{0.02cm}
\begin{picture}(340,480)
\setlength{\unitlength}{0.02cm} \thicklines

\put(200,470){(a)} \put(40,450){$c_{\text{min}}$}
\put(40,320){$c_{\text{max}}$} \put(40,440){\vector(4,0){300}}
\put(40,340){\vector(4,0){300}} \put(40,340){\vector(3,1){300}}
\put(40,440){\vector(3,-1){300}}
\put(340,435){\framebox{$\hat{\cal{O}}_S$}}
\put(340,335){\framebox{$\hat{\cal{O}}_T$}}
\put(220,320){$3(1+c_{\text{max}})/4$}
\put(220,450){$(1-3c_{\text{min}})/4$} \put(270,
370){$3(1+c_{\text{min}})/4$}
\put(270,400){$(1-3c_{\text{max}})/4$}

\put(200,280){$(b)$}

\put(40,270){\mbox{$|\Psi_-\rangle$}}
\put(40,180){\mbox{$|\updownarrow_A\updownarrow_B\rangle$}}
\put(40,260){\vector(4,0){300}} \put(40,200){\vector(4,0){300}}
\put(340,255){\framebox{$\hat{\cal{O}}_S$}}
\put(340,195){\framebox{$\hat{\cal{O}}_T$}} \put(250,180){1}
\put(250,265){1}

\put(200,140){(c)}

\put(40,130){\mbox{$|\updownarrow_A\leftrightarrow_B\rangle$}}
\put(40,0){\mbox{$|\updownarrow_A\updownarrow_B\rangle$}}
\put(40,120){\vector(4,0){300}} \put(40,20){\vector(4,0){300}}
\put(40,120){\vector(3,-1){300}}
\put(340,115){\framebox{$\hat{\cal{O}}_S$}}
\put(340,15){\framebox{$\hat{\cal{O}}_T$}} \put(250,0){1}
\put(250,55){1/2} \put(250,105){1/2}

\end{picture}
\caption{Depiction of the outcomes of operator measurements
$\hat{\cal O}_S$ and $\hat{\cal O}_T$. The general case is shown
in (a). For perfectly correlated noise, when the full range of
allowed entangled states is employed, perfect distinguishability
between the two inputs is possible as shown in (b). In the
restricted separable states only regime, the diagram reduces to
that shown in (c) and the emerging states are unable to be
distinguished unambiguously.} \label{fig:arrows}
\end{figure}

In order to complete the description of the communication
protocol, we need to specify the measurement applied to the states
emerging from the channel. This task can be decomposed into two
steps. The first one is the determination of the total number of
photons contained in the two slots and it can in principle be
accomplished by a collective quantum non-demolition measurement
\cite{QND} on all the modes involved that would determine the
total photon number without destroying coherence between the
modes. Depending on the outcome, the second step needs to be
either finding the temporal slot occupied by a photon in the
one-photon subspace which can be realized by direct temporally
resolved detection, or discriminating between the states used to
encode information in the two-photon subspace. It is easy to see
that this discrimination takes a simple form in the case of
perfectly correlated noise and entangled input states: we need to
determine whether the received states belong to the singlet or the
triplet subspace, which corresponds to a two-element projective
measurement:
\begin{eqnarray}
\hat{\cal{O}}_S & = & |\Psi_- \rangle \langle \Psi_-| \nonumber \\
\hat{\cal{O}}_T & = & \hat{\openone} - |\Psi_- \rangle \langle
\Psi_-|
\end{eqnarray}
It turns out that the same measurement saturates the Holevo bound
also in the general case of any value of the parameter $\eta$ with
either entangled or separable input states. In
Fig.~\ref{fig:arrows}(a) we depict conditional probabilities of
obtaining the singlet or the triplet outcomes for a two-element
input ensemble characterized by Werner parameters $c_{\text{min}}$
and $c_{\text{max}}$. A lengthy but straightforward calculation
shows that if we take as the input probabilities the values
discussed in the preceding section, the mutual information is
given exactly by the right hand side of Eq.~(\ref{Eq:chi(2)le}).
Thus the described procedure indeed maximizes the channel capacity
in the two-photon subspace.

It is instructive to compare the above diagram for optimal
entangled and separable input ensembles in the case of perfect
correlations $\eta=1$. For the optimal entangled ensemble, shown
in Fig.~\ref{fig:arrows}(b) we can distinguish perfectly between
the two inputs as they belong to orthogonal subspaces even after
the transmission. For the separable ensemble, the emerging states
can no longer be perfectly discriminated as seen in
Fig.~\ref{fig:arrows}(c).

\begin{figure}
\epsfig{width=3in,file=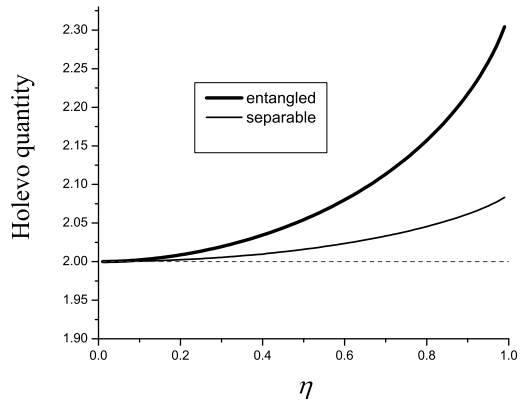} \caption{Graph showing plot of
$\chi$ versus $\eta$. The channel capacity for the general case
where entangled states are used is significantly greater than for
the restricted case where only separable states are employed. The
dashed line is the channel capacity when the polarization degree
of freedom is not used at all.} \label{fig:holevoplot}
\end{figure}

The complete channel capacity obtained by combining
Eq.~(\ref{Eq:ChiSum}) with the results of
Sec.~\ref{Sec:ChannelCapacity} is shown as a function of $\eta$ in
Fig.~\ref{fig:holevoplot}. It is seen that using an entangled
input ensemble gives a clear advantage over the separable states
over the complete range of the correlation parameter $\eta$.

We note that the measurement discriminating between the singlet
and the triplet subspaces can be implemented using the
Braunstein-Mann scheme based on linear optics
\cite{BrauMannPRA95}, as we do not have to distinguish between all
four Bell states. After overlapping temporally the received
photons and interfering them on a 50:50 non-polarizing beam
splitter, their detection in the same output port corresponds to a
projection onto the triplet subspace, whereas measuring them in
the separate output ports of the beam splitter identifies the
singlet state.

\section{Conclusions}
We have introduced a model of a communication channel with
correlated noise motivated by random birefringence fluctuations in
a fiber optic link. Within this model, we have demonstrated that
introducing quantum correlations between consecutive uses of the
channel increases its capacity. This demonstrates how specifically
quantum phenomena such as entanglement can be helpful in the task
of transferring classical information. Making use of entanglement
requires more complex preparation procedures that provide joint
input states extending over a number of temporal slots. A related
question is the role of collective quantum measurements on the
output of the channel rather than detecting radiation in each of
the slots individually and combining classical outcomes of
separate measurements.

The action of the channel has been defined in terms of
transformations of the bosonic field operators. This opens up a
route towards interesting generalizations of the present work, for
example including arbitrary multiphoton states. Another direction
would be extending the model to an arbitrary number of temporal
slots rather than just allowing for correlations between pairs of
consecutive slots as in our example. It is easy to give a simple
protocol showing that in this case the channel capacity can be
enhanced even further. Suppose that the sender generates a train
of zero- and one-photon states with the same probabilities equal
to one half. The first time she is to transmit a photon, she sends
half of maximally entangled pair. In the second instance when a
one photon should be transmitted, she sends the remaining member
of the pair transforming it in such a way that the joint
two-photon polarization state belongs either to the singlet or the
triplet subspace. The receiver implements a
polarization-independent quantum non-demolition measurement on
each temporal slot. When a photon is detected, it needs to be
stored until the arrival of the second member of a pair, when the
discrimination between the singlet and the triplet subspaces can
be performed with the help of a joint measurement. If the
fluctuations in random birefringence can be neglected over the
temporal separation between the photons in a pair, this procedure
allows one to encode one extra bit of information into each pair
of transmitted photons. This gives the average channel capacity
equal to $2.5$ per a pair of temporal slots, enhancing further the
optimal value shown in Fig.~\ref{fig:holevoplot}.

\section*{ACKNOWLEDGEMENTS}
This research was supported by EPSRC and Polish KBN.

\printfigures

\end{document}